\documentclass[twocolumn]{aastex6}
\usepackage{amsmath,amssymb}
\usepackage{hyperref} %% To make sure backref is disabled
\usepackage{ot1patch}

\newcommand{\pir}{{\mathrm \pi}}

\newcommand{\itu}{1}
\newcommand{\sabanci}{2}

\begin{document}

\title[ULX as Super-critical Propellers]
{Ultra-luminous X-Ray Sources as Super-critical Propellers}

\author{M.~Hakan Erkut\altaffilmark{\itu},
	   K.~Yavuz Ek\c{s}i\altaffilmark{\itu},
	   M.~Ali Alpar\altaffilmark{\sabanci}
	  }

\affil{\altaffilmark{\itu} Istanbul Technical University,  Faculty  of Science  and  Letters,  Physics Engineering  Department,
  34469,  Istanbul, Turkey}
\affil{\altaffilmark{\sabanci} Sabanc\i\ University, Orhanl\i\ Tuzla, 34956 {\.I}stanbul, Turkey}
%\affil{\altaffilmark{\fg}  Feza G{\"u}rsey Center for Physics and Mathematics, Bo{\u g}azi{\c c}i University, 34684, {\c C}engelk{\" o}y, Istanbul, Turkey}

\begin{abstract}

We study the evolution of newborn neutron stars in high-mass X-ray binaries interacting with a wind-fed super-Eddington disk. The inner disk is regularized to a radiation-dominated quasi-spherical configuration for which we calculate the inner radius of the disk, the total luminosity of the system and the torque acting on the neutron star accordingly, following the evolution of the system through the ejector and early propeller stages. We find that the systems with $B \gtrsim 10^{13}$~G pass through a short ($\sim 20\,{\rm yr}$) ejector stage appearing as supernova impostors followed by a  propeller stage  lasting $\sim 10^3\,{\rm yr}$. In the super-critical propeller stage the system is still bright ($L\sim 10^{40}\,{\rm erg\, s^{-1}}$) due to the spindown power and therefore appears as an ultra-luminous X-ray source (ULX). The system evolves into pulsating ULX (PULX) when the neutron star spins down to a period ($P\sim 1$~s) allowing for accretion onto its surface to commence. Systems with lower magnetic fields, $B \sim 10^{11}$~G, pass through a long ($10^5\,{\rm yr}$) super-critical propeller stage with luminosities similar to those of the ultra-luminous super-soft sources (ULS), $L \lesssim 10^{40}\,{\rm erg\, s^{-1}}$. The equilibrium periods of these systems in the accretion stage is about $10\,{\rm ms}$, which is much smaller than the typical period range of PULX observed to date.  Such systems could have a larger population, but their pulsations would be elusive due to the smaller size of the magnetosphere. Our results suggest that the ULS and some nonpulsating ULX  are rapidly spinning and highly magnetized young neutron stars at the super-critical propeller stage.

\end{abstract}


\label{firstpage}

\section{Introduction} \label{intro}

The recent detection of pulsations from four ultra-luminous X-ray sources \citep[ULX;][]{bac+14, isr+17a, isr+17b, fur+16, carp+18} not only showed that a large fraction of these objects could be hosting neutron stars \citep{sha15, wik+15, wik+17, kin+17, mid17} accreting matter from a companion object but also that the stellar-mass objects could exceed the classical Eddington limit. 

Another subclass of ULX are the ultra-luminous super-soft sources (ULS) characterized by very soft X-ray spectra $k_{\rm B} T = 0.05 - 0.2$~keV \citep{dis03,fab+03,kon03} whereas the conventional ULX have a better part of their luminosity above $1$~keV. ULX are possibly a heterogeneous class, PULX and ULS' forming subclasses.
It is argued that the differences between conventional ULX and ULS' simply arise from our viewing angle,
ULS being observed at high inclination angles (edge on) so that a thicker layer of material is obscuring the central engine \citep{kyl93,pou+07,fen+16,urq16,pin+17b}.

X-ray spectra of PULX are similar to those of most ULX \citep{pint+17}. Although it is more likely that the compact objects in most ULX are neutron stars, the lack of pulsations is addressed by the likely presence of an optically thick envelope that smears out the pulsations \citep{eks+15,mus+17}. Recently, it was shown by \citet{tsy+16} that the first detected PULX, M82 X-2, has a bimodal luminosity distribution likely because the system occasionally enters into a propeller stage \citep{ill75} when the matter cannot accrete onto the star due to centrifugal barrier. It is then natural to think that some ULX could be systems at an early evolutionary epoch in which the neutron star is 
rotating much faster than a critical rotation rate and is spinning down rapidly, far from spin equilibrium.
In the super-critical (super-Eddington) mass influx regime expected for the ultra-luminous sources
the propeller can easily facilitate outflows and winds from the disk \citep{lov+99}.

In this paper, we advance the view that a fraction of the ULX/ULS population are strongly magnetized $B\sim 10^{11}-10^{13}\,{\rm G}$ neutron stars at the super-critical propeller stage \citep{lip82,min+91,lip99}. In this picture, the spindown energy transferred to the wind-fed disk is the main source of energy at initial stages. Given the evidence for the presence of disks and optically thick outflows, it is likely that some of these objects are  spinning down under propeller torques from quasi-spherical wind-fed disks. In our picture, ULX/ULS systems with $B \gtrsim 10^{13}\,{\rm G}$, (a system with the same properties is seen as a ULX or ULS, depending on the viewing angle) are progenitors of PULX i.e.\ they would become PULX when the neutron star slows down sufficiently. Mass flow toward the neutron star proceeds by wind at the earliest stages of evolution associated with ULX and ULS depending on the dipole magnetic field of the neutron star, but at some stage Roche-lobe overflow commences. This view is consistent with the recent understanding that any ULX system represents a short-lived phase in the life of a binary system \citep{kin+01,wik+15}.

The structure of this paper is as follows: In Section~\ref{sec:prop}, we introduce the basic concepts and equations to derive the spindown torque on the neutron star acting as a supercritical propeller and the luminosity of the disk around this propeller. We present the results of our analysis in Section~\ref{result}. In Section~\ref{disc}, we discuss the astrophysical implications of ULXs being super-critical propellers.

\section{The Super-critical Propeller} 
\label{sec:prop}

The condition for disk formation is that the specific angular momentum of stellar wind matter be larger than the specific angular momentum of matter in Keplerian orbit at the magnetopause \citep{Iben+95,Lu+11}. This requires the relative velocities between the neutron star and the wind to be smaller than those expected for radiatively driven winds. Smaller wind velocities can be due to ionization by the X-ray source and wind flows concentrated toward the neutron star if the optical companion is close to filling its Roche lobe \citep{Livio+86}. Even if the wind matter falls quasi-radially toward the neutron star and thus carries little net angular momentum, a disk can form at the rotation equator following the transfer of angular momentum from the neutron-star magnetosphere to the infalling gas \citep{Anzer+87}. In what follows, we assume that the mass donor transfers matter at supercritical rates $\dot{M}_0>\dot{M}_{\rm E}\equiv L_{\rm E}/\epsilon c^2$, throughout the wind-fed disk around the newborn rapidly rotating neutron star in some ULX/ULS systems.
Here, $L_{\rm E}=4\pir G M m_{\rm p} c /\sigma_{\rm T}$ is the Eddington luminosity, $\epsilon \simeq 0.1$ is the efficiency of gravitational energy release, $m_{\rm p}$ is the proton mass, $c$ is the speed of light, and $\sigma_{\rm T}$ is the Thomson cross-section of the electron.

\subsection{Basic concepts}

The propeller stage is realized when the neutron star rotates so fast that mass cannot be accreted onto the star due to the centrifugal barrier \citep{ill75}. This condition is satisfied when the inner radius of the disk, $R_{\rm in}$, is greater than the corotation radius, $R_{\rm co}=(GM/\Omega_{\ast}^2)^{1/3}$, 
where $M$ is the mass and  $\Omega_{\ast}$ is the angular rotation frequency of the neutron star  \citep{lov+99}.
The system is in the super-critical propeller stage \citep{lip87} if matter is transferred from the donor at a super-critical rate ($\dot{M}_0>\dot{M}_{\rm E}$)
while the corotation radius remains smaller than the
inner radius of the disk because of the rapid rotation of the neutron star.

The super-critical mass transfer within the disk leads to the spherization of the disk within a critical radius,
\begin{equation}
R_{\rm sp} = \frac{27\epsilon \sigma_{\rm T} \dot{M}_0 }{8\pi m_{\rm p} c} \simeq 1.43\times 10^9~{\rm cm}\,\, \epsilon \left( \frac{\dot{M}_0}{10^{20}\,{\rm g\,s^{-1}}} \right),
\end{equation}
determined by $L(R>R_{\rm sp})=27\epsilon GM\dot{M}_0/2R_{\rm sp}=L_{\rm E}$ \citep{sha73}. The flow regulates itself so that some of the matter within the spherization radius is ejected from the system with a radiation-dominated outflow
\begin{equation}
\dot{M} =
\begin{cases}
\dot{M}_0\left(R/R_{\rm sp}\right), \quad & \mbox{for } R<R_{\rm sp}; \\
\dot{M}_0, \quad & \mbox{for } R>R_{\rm sp}.
\end{cases}
\label{mdot}
\end{equation}
\citep{sha73}. Accordingly, the mass flux within the disk is regulated not to exceed the Eddington limit too much, but only logarithmically:
$L \simeq L_{\rm E} [1 + \ln (\dot{M}_0 / \dot{M}_{\rm E})]$
\citep{sha73}.

\subsection{Angular Momentum Loss From the System} \label{amoloss}

A young strongly magnetized neutron star enshrouded in super-critical flow would spindown at a very high rate.
We assume that the angular momentum also is lost with the outflows. For consistency with \autoref{mdot} we write
\begin{equation}
\dot{J} =
\begin{cases}
  \dot{J}_0(R/R_{\rm sp})^{3/2}, \quad & \mbox{for } R<R_{\rm sp}; \\
\dot{J}_0, \quad & \mbox{for } R>R_{\rm sp}.
\end{cases}
\label{jdot}
\end{equation}
Here,
\begin{equation}
\dot{J}_0=\dot{M}R^2\Omega+2\pi R^3\eta \frac{d\Omega}{dR} \label{jdotzero}
\end{equation}
is the angular momentum flux through the radius $R>R_{\rm sp}$ of the disk whose radial size is limited by $R_0$ and $\eta$ is the vertically integrated dynamical viscosity. Using \autoref{mdot} for $R>R_{\rm sp}$ and $\Omega=\Omega_{\rm K}\left(R\right)=\sqrt{GM/R^3}$, we can write
\begin{equation}
\dot{M}_0\sqrt{GMR}\left[1-f\left(R\right)\right]=\dot{J}_0, \label{dimang}
\end{equation}
where
\begin{equation}
f\left(R\right)=\frac{3\pi \eta}{\dot{M}} \equiv 1-\zeta \sqrt{\frac{R_{\rm sp}}{R}} \label{fofr}
\end{equation}
with $\dot{J}_0=\zeta \dot{M}_0\sqrt{GMR_{\rm sp}}$ and $\zeta$ characterizes the dimensionless torque (see~\autoref{totmom}).

\subsection{Propeller Regime}

In the quiescent disk solution \citep{sun77}, sometimes employed for describing the propeller regime \citep{dan10,ozs+14}, the mass flux in the disk is zero so that the material torque $\dot{M}R^2 \Omega$ vanishes and the viscous stress is alone to balance the magnetic stress at the inner rim of the disk. 
For such systems the usual Alfv\'en radius, $r_{\rm A} \propto \dot{M}^{-2/7}$ becomes irrelevant as it can predict an inner disk radius greater than the light cylinder radius \citep{ozs+14}.  
For super-critical propellers, the mass flux is not totally zero throughout the disk, but still it is reduced heavily at the inner rim so that the inner radius of the disk is to be found by the balance of viscous and magnetic stresses. 
Being independent of the specific regime (accretion/propeller), the angular momentum balance near the inner disk radius can be expressed as
\begin{equation}
\dot{J}\left(R_{\rm in}\right)-\dot{J}\left(R_{\rm in}-\Delta R\right)=  - \int_{R_{\rm in}-\Delta R}^{R_{\rm in}} B_{\phi}^{+} B_z R^2\, {\rm d}R,
\label{balance}
\end{equation}
where $\dot{J}$ is the sum of the material and viscous stresses, $B_z\simeq -\mu/R^3$ is the poloidal magnetic field of stellar origin, $B_{\phi}^{+}=\gamma_\phi B_z$ is the toroidal magnetic field above the surface of the disk, and $\gamma_\phi$ is the azimuthal pitch factor of order unity.
The right-hand side of this equation is of the form $ \mu^2\delta/R_{\rm in}^3$, where $\delta \equiv \Delta R / R_{\rm in}$ is the relative width of the coupled domain (boundary region) between the disk and the magnetosphere.
 
\subsubsection{Inner Disk Radius and Spindown Torque} \label{prop}

In the propeller regime, material stresses are negligible at $R_{\rm in}$ and the surface density of the disk matter vanishes just inside the innermost disk radius due to the efficient depletion of matter propelled out by the rapidly rotating magnetosphere, i.e., $\eta \left(R_{\rm in}-\Delta R\right)=0$. This further requires that $\dot{J}\left(R_{\rm in}-\Delta R\right)$ in \autoref{balance} also vanishes as $\dot{J}\propto \eta$. Using \autoref{jdot}  with $\dot{J}_0=\zeta \dot{M}_0\sqrt{GMR_{\rm sp}}$ (see Section~\ref{amoloss}) and $\dot{M}_{\rm in}=\dot{M}_0\left(R_{\rm in}/R_{\rm sp}\right)$ for the super-critical regime in \autoref{mdot}, we find
\begin{equation}
\dot{J}\left(R_{\rm in}\right)=\zeta \dot{M}_{\rm in}\sqrt{GMR_{\rm in}}. \label{totmom}
\end{equation}
As $\dot{J}\left(R_{\rm in}\right)$ represents the net torque acting on the neutron star, the integration constant $\zeta$ can be identified to be the dimensionless torque.  In general, the dimensionless torque, $n$, is a function of the fastness parameter, $\omega_* \equiv \Omega_{\ast}/\Omega_{\rm K}(R_{\rm in})$. We, therefore, write the integration constant as $\zeta=n\left(\omega_*\right)$. The azimuthal pitch is expected to be proportional to the shear between the magnetosphere and the inner disk matter, i.e., $\gamma_\phi \propto \Omega_*-\Omega\left(R_{\rm in}\right)$. Accordingly, we choose the dependence of the azimuthal pitch on the fastness parameter as $\gamma_\phi \left(\omega_*\right)=\gamma_{\rm p}\left(\omega_*-\omega_{\rm p}\right)$ for the propeller regime. Here, $\omega_{\rm p}=\Omega\left(R_{\rm in}\right)/\Omega_{\rm K}\left(R_{\rm in}\right)$ and $\gamma_{\rm p}$ are constants of order unity. Next, we substitute \autoref{totmom} into \autoref{balance} and solve for the innermost disk radius,
\begin{equation}
R_{\rm in}=\left(\frac{\gamma_{\rm p}\omega_{\rm p} \mu^2 R_{\rm sp}\delta}{n_0\dot{M}_0\sqrt{GM}}\right)^{2/9}, \label{rinpro}
\end{equation}
for the super-critical propeller regime $\left(R_{\rm co}<R_{\rm in}<R_{\rm sp}\right)$ assuming that
\begin{equation}
\zeta = n\left(\omega_*\right)=n_0\left(1-\frac{\omega_*}{\omega_{\rm p}}\right) \label{dimtrq}
\end{equation}
in accordance with the dependence of $\gamma_\phi$ on $\omega_*$. The propeller (spindown) torque acting on the neutron star can therefore be written as
\begin{equation}
N_{\rm p}=n_0\left(1-\frac{\omega_*}{\omega_{\rm p}}\right)\dot{M}_{\rm in}\sqrt{GMR_{\rm in}}, \label{ptrq}
\end{equation}
where $n_0$ is another proportionality constant of order unity and $\omega_*>\omega_{\rm p}$.

\subsubsection{Luminosity of the Disk with Outflows}

The energy budget of propeller systems involves the gravitational potential energy released, $L_{\rm G}$, the spindown energy of the neutron star, $L_{\rm sd}$, and the kinetic
energy taken away with the outflowing disk matter, $L_{\rm out}<0$. We express the total luminosity as
\begin{equation}
L_{\rm tot}=L_{\rm{G}}+L_{\rm{sd}}+L_{\rm{out}}. \label{dsklum}
\end{equation}
In the subcritical case, this can be written as
$L_{\rm tot} = GM \dot{M}/R_{\rm in} - I \Omega_{\ast} \dot{\Omega}_{\ast} -\frac12 \dot{M}_{\rm out} v_{\rm out}^2$ \citep[see, e.g.][]{eks+05}. 
Super-critical propellers, however, need further care as the gravitational energy released outside and inside the spherization radius require separate treatment.

For a disk with super-critical mass transfer rates, the rate of potential energy release, $L_{\rm{G}}=\int_{R_{\rm in}}^{R_0}GM\dot{M}\, {\rm d}R/R^2$, can be found as
\begin{equation}
L_{\rm{G}}= \frac{2}{27\epsilon} \left[ \ln{\left(\frac{R_{\rm sp}}{R_{\rm in}}\right)} + 1-\frac{R_{\rm sp}}{R_0} \right] L_{\rm E}
\label{lumgrav}
\end{equation}
where we employed \autoref{mdot}.

 \begin{figure*}
\includegraphics[width=0.49\linewidth]{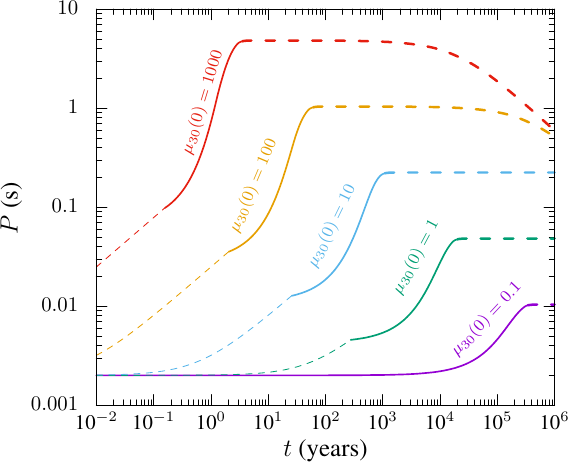}
\includegraphics[width=0.49\linewidth]{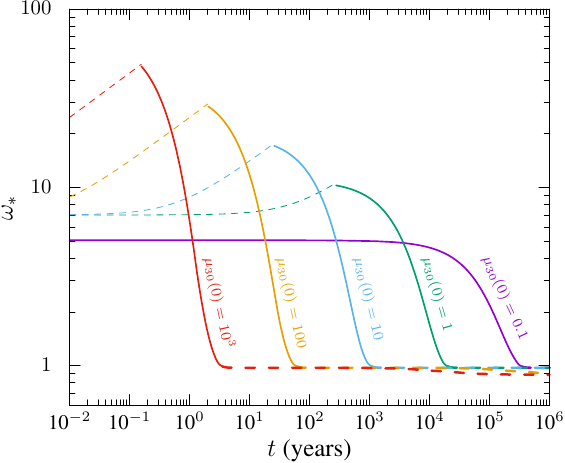}
\caption{Evolution of the spin period (left panel) and the fastness parameter (right panel) of neutron stars for a variety of initial magnetic fields. The dashed thin lines correspond to the ejector stage, the solid lines correspond to the propeller stage we focus in this work, and the thick dashed lines correspond to the late accretion stage. In all simulations the initial period of the neutron star is taken as $P_0 = 2$~ms, mass flux of $\dot{M}_0= 4\times 10^{20}\, {\rm g\, s^{-1}}$, and $R_0=10^{11}\,{\rm cm}$. The dipole magnetic moments greater than $\mu = 10^{31}$~G~cm$^3$ are assumed to decay according to the scenario B of \citet{col+00}.}
\label{fig:period}
\end{figure*}

The energy loss rate due to outflows at the inner region is
\begin{equation}
L_{\rm out}=\frac{1}{2}\int_{R<R_{\rm sp}}v_{\rm out}^2 \, {\rm d}\dot{M}_{\rm out} \label{lout}
\end{equation}
Using  ${\rm d}\dot{M}_{\rm out}=-{\rm d}\dot{M}$ and
\begin{equation}
v_{\rm out}^2\left(R\right)=\frac{2GM}{R}\left(\frac{L_{\rm tot}}{L_{\rm E}}-1\right) \label{vout}
\end{equation}
\citep{sha73} for $L\geq L_{\rm E}$, \autoref{lout} can be evaluated as
\begin{equation}
L_{\rm out}=-\frac{1}{27\epsilon}\left(\frac{L_{\rm tot}}{L_{\rm E}}-1\right) \ln{\left(\frac{R_{\rm sp}}{R_{\rm in}}\right)^2} L_{\rm E} \label{outlum}
\end{equation}
where we refer to \autoref{mdot} in the last step.

The spindown power in \autoref{dsklum} can be written as
\begin{equation}
L_{\rm sd}=-N_{\rm p}\Omega_{\ast}=-n\left(\omega_*\right)\dot{M}_{\rm in}\sqrt{GMR_{\rm in}}\Omega_{\ast}. \label{sdpwr}
\end{equation}
Using $\omega_*$, $R_{\rm sp}$, and \autoref{mdot}, this expression can be further simplified as
\begin{equation}
L_{\rm sd}=-\frac{2}{27\epsilon}\omega_{\ast} n\left(\omega_{\ast}\right)L_{\rm E}. \label{lumsd}
\end{equation}
The luminosity of the disk interacting with a neutron star at the super-critical propeller regime is finally obtained as
\begin{equation}
\cfrac{L_{\rm tot}}{L_{\rm E}}=\cfrac{\ln\left(\cfrac{R_{\rm sp}}{R_{\rm in}}\right)^2+\left(1-\cfrac{R_{\rm sp}}{R_0}\right)-\omega_{\ast} n(\omega_{\ast})}{\cfrac{27\epsilon}{2}+\ln\left(\cfrac{R_{\rm sp}}{R_{\rm in}}\right)}. \label{finlum}
\end{equation}
In the presence of beaming $b<1$, the luminosity would appear even larger: $L_{\rm obs} = L_{\rm tot}/b$.

\begin{figure*}
\centering
\includegraphics[width=0.48\linewidth]{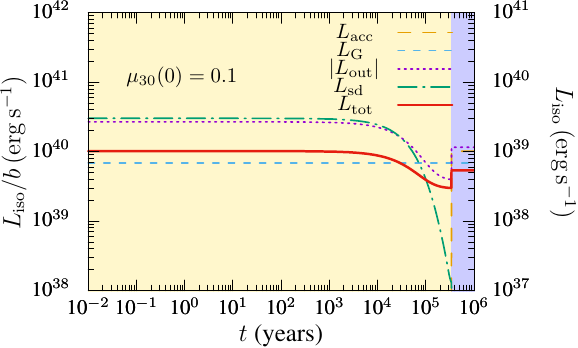}
\hspace{10pt}
\includegraphics[width=0.48\linewidth]{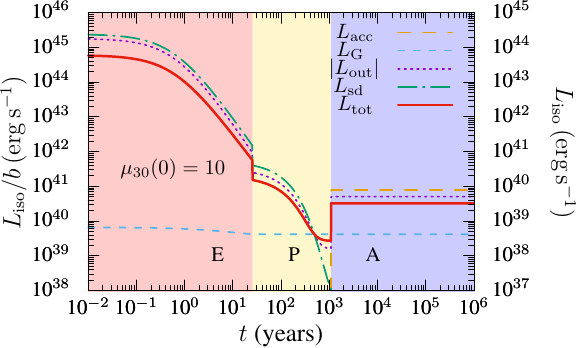}

\caption{Evolution of the luminosity components of neutron stars for a initial magnetic dipole moment of $\mu_{30}(0)=0.1$ (left panel) 
and for $\mu_{30}(0)=10$ (right panel). As in Figure 1 the initial period of the neutron star is taken as $P_0 = 2$~ms, 
mass flux of $\dot{M}_0= 4\times 10^{20}\, {\rm g\, s^{-1}}$, and $R_0=10^{11}\,{\rm cm}$. The right vertical axis represents the luminosity for isotropic emission. The observed luminosity given on the left vertical axis is $L_{\rm obs}=L_{\rm iso}/b$, where $b$ is assumed to be $0.1$. The luminosity enhancement at the late stage ($t \gtrsim 10^5\,{\rm yr}$ for the left panel and $t \gtrsim 10^3\,{\rm yr}$ for the right panel) corresponds to the super-critical accretion regime where the source might be observed as a PULX. The super-critical propeller regime we focus on in this work corresponds to $t \lesssim 10^5\,{\rm yr}$ for the left panel and $t \sim 30-1000\,{\rm yr}$ for the right panel. In this regime, the source would appear as a nonpulsating ULX. The very early (ejector) phase on the right panel ($t \lesssim 10\,{\rm yr}$) may appear as a supernova impostor. E, P and A labeling differently colored shaded regions denote the ejector, propeller and accretion stages, respectively.}

\label{fig:luminosity}
\end{figure*}

The total luminosity in the very late super-critical accretion stage is  $L_{\rm tot}=L_{\rm{G}}+L_{\rm{out}} + L_{\rm acc}$, where $L_{\rm acc}=GM\dot{M}_{\rm in}/R_*$ is the accretion luminosity for a neutron star of radius $R_*$.

\section{Results} \label{result}

We solved the torque equation, $I \dot{\Omega}_{\ast}=N$, by a standard numerical scheme to find the spin and luminosity evolution of neutron stars of moment of inertia $I$, under super-critical mass inflow.

For the initial magnetic field strengths in the magnetar range, i.e.\ for $\mu(0)>10^{31}$~G~cm$^3$, we allowed the field decay and employed the mechanism B in \citet{col+00} as an illustrative example. In the slowest field-decay scenario, mechanism A in \citet{col+00}, the field strength remains constant throughout the evolutionary timescale leaving the maximum equilibrium period unchanged. For the fastest field-decay scenario, mechanism C in \citet{col+00}, the maximum equilibrium period will remain unchanged whereas the asymptotic value of the equilibrium period converges to the equilibrium period attained by $\mu_{30}(0)=10$ in $\sim 10^6\,{\rm yr}$.

In this work,  we focus on the propeller regime and we did not include accretion induced field decay. Our results for the accretion stage would change if accretion induced field decay is considered; e.g.\ initial magnetic dipole moments even stronger than $\mu_{30}(0) = 10$ would be required to obtain PULX periods. The accretion regime is beyond the scope of this paper and will be studied thoroughly in a subsequent work for different accretion induced field-decay scenarios with additional set of parameters to be scanned for addressing observational properties of PULXs.

For a  neutron star of mass $M=1.4\, M_{\odot}$ and radius $R_*=10$~km we assumed the initial period $P_0 = 2$~ms, mass flux $\dot{M}_0= 4\times 10^{20}\, {\rm g\, s^{-1}}$, and the outer disk radius $R_0=10^{11}\,{\rm cm}$.
We also assumed that $\epsilon=0.1$, $\delta = 0.01$, $n_0=1$, $\omega_p=0.9$, and $\gamma_p=0.8$.
We obtained results for a range of initial magnetic moments $\mu_{30}(0)=0.01$, $\mu_{30}(0)=0.1$, $\mu_{30}(0)=1$, $\mu_{30}(0)=10$, $\mu_{30}(0)=100$, and $\mu_{30}(0)=1000$, where $\mu_{30}(0)\equiv \mu(t=0) / 10^{30}\,{\rm G\,cm^3}$. The period and fastness parameter evolution of these systems is shown in \autoref{fig:period}. We see that systems with $\mu_{30}(0)<1$ start in the propeller stage whereas those with $\mu_{30}\geq 1$ start as ejectors. In the ejector stage, the inflowing material does not even reach the light cylinder and the neutron star spins down as an isolated rotating dipole \citep{lip92}.

The evolution of each luminosity component is shown in \autoref{fig:luminosity} for  neutron stars with different magnetic fields. The left and right panels show the long-term early-stage evolutions of each luminosity component discussed in the previous section. In line with the period evolution of the same systems with $\mu_{30}=10$ and $\mu_{30}=1$ in \autoref{fig:period}, the neutron stars appear as ULXs for an early period of $10^3-10^4\, {\rm yr}$ during which they attain their equilibrium period.

As seen from \autoref{fig:period}, only systems in the range $100>\mu_{30} >1000$ have equilibrium periods in the observed range of $\sim 1\,{\rm s}$. The equilibrium period can be guessed from the torque expression in \autoref{ptrq}. For $N_{\rm p}=0$, $\omega_*=\omega_{\rm p}\simeq1$ must be satisfied. This condition yields the equilibrium period, $P_{\rm eq}\simeq 0.21 \, {\rm s}\,\mu_{31}^{2/3}\delta_{0.01}^{1/3}$. Note, however, that longer periods are also possible for wider zones of magnetosphere-disk interaction. In the accretion regime, widths of transition zones as large as $0.6$ can be realized \citep[see, e.g.][]{Erkut04}. For $\delta=0.6$ and $\mu_{31}=1$, $P_{\rm eq}\simeq 0.8\, {\rm s}$.

\section{Discussion and Conclusions}
\label{disc}

We have proposed that some ULX and ULS sources host neutron stars in the super-critical propeller stage corresponding to super-Eddington mass inflow rates and with the radiation pressure in the inner disk sustaining a quasi-spherical geometrically thick structure.
This leads to a natural mechanism for the optically thick winds with velocity $v \sim 0.2 c$ launched by radiative and centrifugal processes.
The X-ray emission is then processed in this optically thick medium and is observed as soft emission at large viewing angles.
The spindown energy released by the slowing down neutron star is the main source of the extreme luminosity of these objects.

We have shown that only those systems with initial magnetic fields $B\gtrsim 10^{13}\,{\rm G}$ are spun down to periods $P\sim 1\,{\rm s}$ matching the observed periods of PULXs confirming the early view \citep{eks+15} that the first discovered PULX, M82 X-2 has such strong magnetic fields. Inclusion of accretion induced magnetic field decay will obviously require even stronger initial field strengths. We have estimated the spindown torque and shown that the spindown power at the early stage when the neutron star is rotating rapidly is sufficient to address the luminosity of ULX and ULS systems.

\subsection{Evolution from Massive Binaries}

We now check that super-Eddington mass inflow rates $\dot{M}_0$ indicated for ULX evolutionary scenarios are indeed provided by binary evolution, for the required duration of
$\sim 10^6\,{\rm yr}$.

In the usual approach to the rotational evolution of neutron stars in massive binaries, prior to the start of accretion, the secular increase of the neutron-star spin period is realized though two successive phases: (i) the ejector phase in which the young neutron star can emit radiation from radio waves to X-rays and (ii) the propeller phase, where the infalling matter of wind from the massive donor is thrown away by the rapidly rotating neutron-star magnetosphere \citep{ill75}. In the propeller phase, the system is assumed to be unobservable or extremely faint in X-rays, because no matter is allowed to accrete onto the neutron-star surface. In the absence of accretion, this assumption can be validated if the spindown power transferred by the neutron-star magnetosphere to the accretion flow remains sufficiently low as for the sub-Eddington mass-inflow rates. The brightest X-ray stage for neutron stars in high-mass X-ray binaries is then anticipated to begin only when the massive component is close to filling its Roche-lobe and starts transferring its mass via an accretion disk \citep{Iben+95}. Once the magnetospheric radius in the inner disk becomes smaller than the corotation radius, the neutron star is said to be X-ray luminous in the so-called accretion phase.

Long before the Roche-lobe overflow occurs, the neutron-star magnetosphere can interact with the wind of the massive companion. In the present evolutionary scheme, we consider the case of a newly born neutron star embedded in the wind of the massive companion. In the earliest stage with neutron-star spin periods of a few milliseconds, the substantial mass transfer to the neutron star can be realized if the donor is an already evolved massive star that is capable of producing dense winds with sufficiently high mass-loss rates. The formation of a helium star--neutron star binary, following an early common-envelope phase in the course of the evolution of two stars of nearly equal initial masses (twin massive binaries), is the best example of a neutron-star birth in a binary where the massive component starts feeding the compact object through powerful winds \citep{Brown95,Dewi+06}. The evolution of such binaries usually ends up with double neutron-star systems soon after the explosion of the helium star.

Cygnus~X-3 is the only known candidate in our galaxy for a neutron star (or a low-mass black hole) accreting matter from a massive helium star \citep[Wolf--Rayet star;][]{Lommen+05,Zdziarski+13}. Cygnus~X-3, albeit luminous in X-rays $\left(L_{\rm X}\simeq 10^{38} \, {\rm erg} \, {\rm s^{-1}}\right)$, is not a ULX. It is, however, possible that Cygnus~X-3 represents the late stage of a ULX  at which the mass transfer rate has already been reduced to $\dot{M}_0 \simeq \dot{M}_{\rm E}$. The evolution of helium stars, such as the one in Cygnus~X-3, is similar to the evolution of Wolf--Rayet (WR) stars with strong mass loss \citep{McClelland16}. The typical range for the mass-loss rates due to winds of massive helium/WR stars is $\sim 10^{-6}-10^{-4} \, M_\odot \, {\rm yr^{-1}}$ \citep{Lommen+05,Crowther07,Zdziarski+13}.

Being hot enough, the single hydrogen-poor WR (helium) stars can produce winds with velocities as high as $\sim 4000\,{\rm km\,s^{-1}}$ \citep{Crowther07}. As revealed by the hydrodynamic atmosphere models for WR stars \citep{Grafener05}, the radiatively driven wind structure consists of two acceleration regions. The region that is close to the wind base of the stellar atmosphere is characterized by an optically thick wind of velocities $v_{\rm w}\lesssim 1000\,{\rm km\,s^{-1}}$. The second region where the wind velocities exceed $1000 \,{\rm km\,s^{-1}}$ extends across the outer part of the wind. In a close binary, such as Cygnus~X-3, the wind velocities can therefore be as low as $1000 \,{\rm km\,s^{-1}}$ even in the absence of any wind-velocity reduction mechanism.

The observed spectrum of the mass-donor star has been reconstructed in a recent hydrodynamical atmosphere model of the high-mass X-ray binary, Vela~X-1 \citep{Sander18,Sander+18}. The wind velocity at the location of the neutron star has been estimated as $\sim 100\,{\rm km\,s^{-1}}$, which is much smaller than the typical value of the wind velocity expected according to the standard approach. 
In the classical Bondi--Hoyle theory, slow wind favors relatively high accretion rates for the wind material (see, e.g., \autoref{mlof}). Even if the wind velocities are not as low as the classical theory suggests, the efficiency for the mass transfer through wind accretion may be much higher in X-ray binaries with WR stars than previously thought as in the case of M101~ULX-1 \citep{liu+13}.

In a high-mass X-ray binary, where the massive donor is X-ray irradiated by the compact X-ray source, the mass transfer can be significantly altered due to the effect of X-ray photoionization of the wind material \citep{Ducci+10}. As revealed by simulations of the stellar wind in X-ray binaries, the X-ray photoionization decelerates the wind matter in the vicinity of the compact object, leads to the formation of an extensive disk, and therefore enhances the overall mass inflow \citep{Cechura15}. The wind velocities of $\sim 200\, {\rm km\,s^{-1}}$ can be realized if the wind driving radiative processes are suppressed by the effect of X-ray photoionization \citep{Sako+02}. As also mentioned in Section~\ref{sec:prop}, the formation of a disk around the newborn neutron star as an ejector (right panel of \autoref{fig:luminosity}) or a propeller (left panel of \autoref{fig:luminosity}) is favored by the reduction of wind velocities due to the X-rays emitted from the young neutron star.

Low wind velocities are not sufficient for the formation of accretion disks around wind-fed compact objects. According to the disk-formation criterion, there is an upper limit for the orbital period of the binary, which depends on the masses of binary components as well in addition to the wind velocity \citep{Lommen+05}. In order to determine the current population of helium star--neutron star (and helium star--black hole) binaries in our galaxy, \citet{Lommen+05} performed a population synthesis and obtained the distribution of core-helium-burning systems in the orbital period versus helium-star mass plane. For a given orbital period and a wind velocity, \citet{Lommen+05} revealed that the disk formation around the neutron star is more likely for massive helium stars with masses $>7M_\odot$ compared to low-mass helium stars with masses $<5M_\odot$, though the latter is much more abundant than the former.

The fraction of the mass-loss rate, $\dot{M}_{\rm w}$, of the massive helium wind supplier captured by the neutron star is given by
\begin{equation}
\frac{\dot{M}_0}{\dot{M}_{\rm w}} \simeq 3.2\times 10^{-3} \, M_{1.4}^{4/3} P_{\rm 1\,d}^{-4/3} \left(1+q\right)^{-2/3} v_{\rm w,1000}^{-4} \label{mlof}
\end{equation}
\citep[see, e.g.,][]{Urpin+98}. Here, $q\equiv M_{\rm D}/M$ is the mass ratio of the helium star--neutron star binary with $M_{\rm D}$ being the mass of the helium donor, $P_{\rm 1\,d}$ is the binary orbital period in units of 1 day, $v_{\rm w,1000}\equiv v_{\rm w}/1000\,{\rm km\,s^{-1}}$ is the wind velocity at the neutron-star location, and $M_{1.4}$ is the neutron-star mass in units of $1.4\, M_\odot$. For the mass-transfer rates of $\dot{M}_0=4\times 10^{20} \, {\rm g\, s^{-1}}\,(\simeq 6\times 10^{-6} \, M_\odot \, {\rm yr}^{-1})$ we employ in the present work, the mass-loss rate for the WR star (massive helium donor) with $M_{\rm D}=15\,M_\odot$ can be estimated using \autoref{mlof} with $v_{\rm w}=200\,{\rm km\,s^{-1}}$ as $\dot{M}_{\rm w}\simeq 1.5\times 10^{-5}\, M_\odot \, {\rm yr}^{-1}$, which is in agreement with the observed mass-loss rates of WR stars \citep{Zdziarski+13}.

The lifetime of the helium-burning phase for a star of mass $\gtrsim 15\, M_\odot$ is $\sim 10^6\,{\rm yr}$ \citep{Hayashi62,Chiosi+78,Salasnich+99}. Following the helium burning, the fusion of the carbon, oxygen, and other heavy elements occurs within $\lesssim 10^4\,{\rm yr}$ before the donor core collapses into a neutron star. According to our evolutionary scenario, a massive helium star--neutron star binary is left behind an early common-envelope phase during the evolution of two massive main-sequence stars. The accretion disk around the neutron star is fed by the helium-star wind transferred with an average mass flux of $\dot{M}_0 \sim 6\times 10^{-6} \, M_\odot \, {\rm yr}^{-1}$ throughout the helium-burning lifetime. Such a mass-transfer rate can be sustained by the mass-loss rate that evolves in time according to \autoref{mlof}. Noting that $\dot{M}_{\rm w}=-dM_{\rm D}/dt$ and using typical values such as $M_{1.4}=1=P_{\rm 1d}$ and $v_{\rm w,1000}=0.2$ for all quantities except $M_{\rm D}$ as a simplifying assumption, it follows from the integration of \autoref{mlof} over $10^6\,{\rm yr}$ that the terminal mass of the helium star toward the end of helium-burning phase (before carbon burning starts) is $\sim 4\, M_\odot $ for an initial helium-star mass of $15\, M_\odot $ (at the onset of helium burning) if mass-inflow rates are super-Eddington with $\dot{M}_0 \sim 6\times 10^{-6} \, M_\odot \, {\rm yr}^{-1}$ as appropriate for ULX. The evolutionary scenario we presently invoke to explain a subgroup of ULXs (such as the ULS and some nonpulsating ULX) cannot, however, account for the observed population of X-ray binaries with low-mass helium stars. Instead, the majority of systems with low-mass helium stars are likely to be the direct outcome of an early common-envelope phase in the course of the evolution of two massive stars \citep{Bhattacharya91}.

The chemical composition of the wind-fed disk around the neutron star is determined by the ingredients of the helium-star envelope. In case all hydrogen is depleted, a helium-rich disk with the same critical rate of mass inflow can be realized provided $\left(L_{\rm E}/\epsilon \right)_{\rm He}=\left(L_{\rm E}/\epsilon \right)_{\rm H}$. The Eddington luminosity for the accretion of the helium-rich matter is twice the Eddington luminosity for the hydrogen-rich gas. The efficiency of the helium-rich disk is then $\epsilon_{\rm He}=2\epsilon_{\rm H}\simeq 0.2$ for $\epsilon_{\rm H}\simeq 0.1$ (Section~\ref{sec:prop}). It is possible to compare the luminosities of the helium- and hydrogen-rich disks assuming the same numerical values for all model parameters except $\epsilon$. Note from \autoref{lumgrav} and \autoref{sdpwr} that $\left(L_{\rm G}\right)_{\rm He}=\left(L_{\rm G}\right)_{\rm H}$ and $\left(L_{\rm sd}\right)_{\rm He}=\left(L_{\rm sd}\right)_{\rm H}$. Using \autoref{finlum}, on the other hand, we find
\begin{equation}
\frac{\left(L_{\rm tot}\right)_{\rm He}}{\left(L_{\rm tot}\right)_{\rm H}}=1+\left[1+\frac{27\epsilon_{\rm H}}{\ln\left(R_{\rm sp}/R_{\rm in}\right)}\right]^{-1}. \label{cmplum}
\end{equation}
As seen from \autoref{cmplum}, $\left(L_{\rm tot}\right)_{\rm He}\simeq\left(L_{\rm tot}\right)_{\rm H}$ for $R_{\rm sp}/R_{\rm in}\simeq 1$, which can be satisfied for sufficiently strong magnetic fields. For magnetic fields as low as $10^{11} \, {\rm G}$ in strength, $R_{\rm sp}/R_{\rm in}\simeq 80$ and $\left(L_{\rm tot}\right)_{\rm He}\simeq 1.6\left(L_{\rm tot}\right)_{\rm H}$. The difference between the helium- and hydrogen-rich disk luminosities is therefore negligible throughout the evolutionary lifetime ($10^6\,{\rm yr}$) as far as the ULX luminosity range is concerned, so that our scenario is not sensitive to composition.

\subsection{Neutron Stars Rather than Black Holes}

Given the abundance of neutron stars over black holes as the outcome of stellar evolution,
it was proposed that the bulk of the ULX population may consist of neutron stars in binary systems rather than accreting stellar-mass black holes \citep{fra+15,sha15,kin+17,mid17,wik+17}. Because of the large mass ratio in a neutron-star high-mass X-ray binary, however, the mass transfer is not stable unlike the black hole systems, where Roche-lobe overflow is stable rendering such sources relatively long-lived as compared to their neutron-star counterparts and thus strong candidates for nonpulsating ULX \citep{Rappaport+05}.

The support in favor of neutron stars being dominant in the ULX population is provided by \citet{pint+17} who show that some of the nonpulsating ULXs they studied exhibit similar X-ray spectra with PULXs. Specifically, 2 of the 12 sources analyzed by \citet{pint+17} show the hard power-law with exponential cutoff component that is likely associated with the accretion column and are thus likely to be accreting sources. The lack of pulsations from these systems could be due to an optically thick medium smearing out the pulsations \citep{eks+15, mus+17}. For the rest of the sources that lack the hard component and have softer spectra, the super-critical propeller regime that we consider in this work cannot be excluded.
ULS' are likely to be super-critical propeller systems with lower magnetic dipole moments, $\mu \sim 10^{29}$~G~cm$^3$ (see the left panel of Figure~\ref{fig:luminosity}) given that the propeller stage of these systems has a longer life-time. Such low-B systems at the accretion stage will have spin periods as small as $P \sim 10$~ms yet may not show pulsations as their magnetosphere enshrouded by super-Eddington accretion is very small.

In the presence of beaming the characteristic spindown time-scale becomes
\begin{equation}
\tau_{\rm c} = 3.1\times 10^6\, {\rm yr}\, b_{0.01}^{-1} L_{40,{\rm iso}}^{-1} P_{2\,{\rm ms}}^{-2} I_{45}
\end{equation}
where $b_{0.01}=b/0.01$ is the beaming fraction in units of $0.01$, $P_{2\,{\rm ms}}=P_0/(2\,{\rm ms})$ and $L_{40,{\rm iso}}$ is the isotropic luminosity in units of $10^{40}$~erg~s$^{-1}$. This shows that at least $10\%$ of the NS population of the ULX/ULS systems that evolve from neutron star--helium star binaries could be in the supercritical propeller stage with duration $\lesssim 10^5\,{\rm yr}$ (\autoref{fig:luminosity}) if the evolutionary timescale is $\sim10^6\,{\rm yr}$.

The first discovered PULX M82 X-2 is known to show bimodal luminosity behavior in the archival data \citep{tsy+16}. The authors interpret the low-luminosity stage as an evidence for the transition of the source to the propeller regime. We find it necessary to emphasize that
this propeller stage is distinct from the early super-Eddington propeller phase we consider
in this work, as the neutron star in this system has already slowed down to $P \sim 1$~s and the spindown power is no longer sufficient for making the object appear ultra-luminous in the observed late propeller stage.

A final note is about the anomalous X-ray pulsar (AXP) 4U~0142$+$61.
A supernova fallback disk is detected in this system by \citet{wan+06} and likely around 1E~2259$+$586 \citep{kap+09}, which also is an AXP.
Such disks were proposed to exist around young pulsars \citep{mic81} at a time when AXPs had not been identified.
Later on they were proposed as an alternative to the magnetar picture \citep{cha+00,alp01} and as an ingredient of magnetars with
strong magnetic fields in multipoles \citep{eks03,ert03}.
Whether the detected disk is  passive \citep{wan+06} or active, as is likely \citep{ert+07}, at the present time, it must have been highly active
at its earliest stages with super-critical mass inflow passing through a super-critical propeller stage \citep{eks03,yan+12} and must have been as bright as an ULX/ULS system.
This implies that some fraction of ULX or ULS systems though not in binary systems could form super-critical propeller systems with the supernova fallback disks
and form the progenitors of AXPs and other classes of young neutron stars.

\section*{Acknowledgments}

M.H.E. acknowledges the post-doctoral research support from the BAP unit of Istanbul Technical University. M.A.A. is a member of the Science Academy (Bilim Akademisi), Turkey.

\bibliographystyle{aasjournal}
\footnotesize{
\bibliography{refs}
}

\label{lastpage}

\end{document}